\documentclass{JHEP3}
\usepackage{graphics}
\usepackage{amsmath}
\usepackage{cite}
\newcommand{\PL}[2]{\, {\rm Li}_{#1}\!\left({#2}\right)}
\newcommand{\hs}[1]{\hspace*{#1 pt}}

\newcommand{\lp}{\lambda_+}
\newcommand{\lm}{\lambda_-}
\newcommand{\loopint}[1]{\int \!\!\! \frac{d^D #1}{\left(2\pi\right)^D}\!}
\newcommand{\ESGamma}{S_{\Gamma}}
\newcommand{\MB}[2]{\hs{-12} \int\limits_{\hs{15}_{ #1 -i \,
\infty}}^{\hs{15}^{ #1 +i\, \infty}} \hs{-15} \frac{d #2}{2\pi i}}

\def\be{\begin{equation}}
\def\ee{\end{equation}}
\def\bea{\begin{eqnarray}}
\def\eea{\end{eqnarray}}
\def\eps{\epsilon}
\def\nnb{\nonumber}
\def\eps{\epsilon}

\preprint{PITHA-09/01\\
SFB/CPP-09-04}
\title{On a two-loop crossed six-line master integral with two massive lines}

\author{T.~Huber\\
Institut f\"{u}r Theoretische Physik E, \\RWTH Aachen University,\\
D-52056 Aachen, Germany\\
\email{thuber@physik.rwth-aachen.de}}

\abstract{We compute the two-loop crossed six-line vertex master integral with two massive lines in dimensional regularisation, and give
the result up to the finite part in $D-4$. We apply the differential equation technique, and focus in particular on the purely
analytical calculation of the boundary condition which we derive from a three-fold Mellin-Barnes representation. We also describe how
the computation of the boundary condition is used to derive three non-trivial relations among harmonic polylogarithms of the sixth root
of unity.
}


\begin{document}

\section{Introduction}\label{sec:intro}

The computation of higher order perturbative corrections to heavy-to-light currents has recently been an active field of research. The
analytical calculation of two-loop QCD corrections to differential semi-leptonic $b \to u $ decays has been carried out simultaneously
by several groups~\cite{Bonciani:2008wf,Asatrian:2008uk,Beneke:2008ei,Bell:2008ws}. However, one of the occurring master integrals,
namely the crossed six-line integral with two massive lines, has not been calculated purely analytically by any of the aforementioned
groups.

In this article we close this gap by reanalysing the master integral in question. We rederive the result through order ${\cal
O}((D-4)^0)$ by purely analytical steps, thereby confirming a suggestion in terms of transcendental constants for the boundary condition
of the finite part in $D-4$. During the course of the calculation we also found three non-trivial relations among harmonic
polylogarithms (HPLs) of the sixth root of unity.

This paper is organised as follows. In section~\ref{sec:def} we define the kinematics and our notation and summarize the final result.
In section~\ref{sec:bc} we elaborate on the boundary condition which we derive from a three-fold Mellin-Barnes (MB) representation. In
section~\ref{sec:hpl} we show how to derive from the MB expressions three non-trivial relations between HPLs of the sixth root of unity.
We conclude in section~\ref{sec:conc}.

\section{Definitions and results}\label{sec:def}
\FIGURE{
\begin{picture}(150,28.5)
\put(75,14.25){\makebox(0,0){\includegraphics{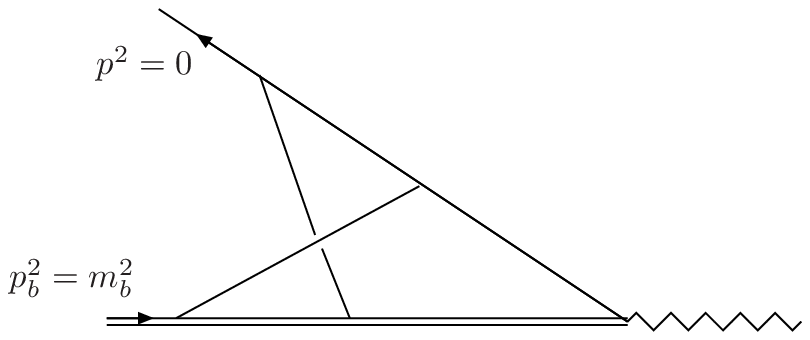}}}
\end{picture}
\caption{The two-loop crossed six-line master integral with two massive (double) lines of mass $m_b$, and four massless (single) lines.
The horizontal line has incoming momentum $p_b$ with $p_b^2=m_b^2$, and the sloped line has outgoing
momentum $p$ with $p^2=0$.}
\label{fig:masters}}
In the following we consider the two-loop crossed six-line master integral depicted in Fig.~\ref{fig:masters}. We work in dimensional
regularisation with $D=4-2\eps$ dimensions which is used to regularise both UV and IR divergences. Moreover, we define
\be
x \equiv \frac{(p_b-p)^2+i\eta}{m_b^2} \; ,
\ee
where $+i\eta$ stems from the $+i\eta$ prescription which we tacitly assume to be included in the propagators.
Defining also the symbols
\be
\int \! \left[dk\right] \equiv \loopint{k} \; , \qquad \ESGamma \equiv \frac{1}{\left(4\pi\right)^{D/2} \Gamma(1-\eps)} \; ,
\ee
our starting expression reads
\bea
I &=& \int \! \left[dk_1\right] \!\int \! \left[dk_2\right] \frac{1}{\left[(k_2+p_b)^2-m_b^2\right] (k_2+p)^2
\left[(k_1+p_b)^2-m_b^2\right] (k_2-k_1+p)^2} \nnb\\
&& \hspace*{74pt}\times \frac{1}{(k_2-k_1)^2 \, k_1^2} \; .
\label{eq:I}
\eea

The integral in question was first given
in~\cite{Bell:2006tz} and subsequently used in~\cite{Bell:2007tv,Bonciani:2008wf,Asatrian:2008uk,Beneke:2008ei,Bell:2006tz,Bell:2008ws,guidorealpart}. Also the
cases of six massless lines~\cite{Gonsalves:1983nq,vanNeerven:1985xr,Kramer:1986sr,Moch:2005id,Gehrmann:2005pd,Moch:2005uq}, as well as
one~\cite{Aglietti:2003yc}, two~\cite{Aglietti:2004ki,Aglietti:2007as}, and four
massive lines~\cite{Bonciani:2003te,Bernreuther:2004ih,Bernreuther:2004th} are available in the literature. In the latter case the
topology has two master integrals. Whereas the purely massless integral reveals a closed form in terms of hypergeometric functions of
unit argument, the cases with massive lines are solved order by order in $\eps$ by means of the differential equation
technique~\cite{Kotikov:1990kg,Kotikov:1991hm,Kotikov:1991pm,Remiddi:1997ny,Caffo:1998du,Caffo:1998yd,Argeri:2007up}.

The $x$-dependence of the integral $I$ is obtained by taking the derivative of $I$ w.r.t.\ $x$ and plugging in the result of a Laporta
reduction~\cite{Laporta:1996mq,Laporta:2001dd,Anastasiou:2004vj}. This
yields a linear combination of the integral $I$ itself and 16 other (fewer-line) master integrals, the latter are known to sufficiently high orders
in $\eps$~\cite{Bell:2006tz}. The solution of the differential equation yields the $x$-dependence except for the boundary condition at
$x=0$.

Contrary to the case of four massive lines~\cite{Bonciani:2003te}, the boundary condition at $x=0$
of the present integral cannot be inferred from the condition that the integral be regular at $x=0$, but has to be calculated explicitly
by means of other techniques. We summarize here our analytical result and postpone the derivation of the boundary condition at $x=0$
to section~\ref{sec:bc}. Through order ${\cal O}((D-4)^0)$ the result is
\bea
I&=& - \ESGamma^2 \, (m_b^2)^{-2-2\eps}
\left\{\frac{c^{(-4)}}{\eps^4}+\frac{c^{(-3)}}{\eps^3}+\frac{c^{(-2)}}{\eps^2}+\frac{c^{(-1)}}{\eps}+c^{(0)} + {\cal O}(\eps)\right\}
\eea
with
\bea
c^{(-4)} &=&\frac{1}{12 \,(1-x)^2} \; , \nnb
\eea
\bea
c^{(-3)} &=&-\frac{\ln(1-x)}{3 \,(1-x)^2} \; , \nnb \\
c^{(-2)} &=&\frac{1}{72 \,(1-x)^2} \left[48 \ln^2(1-x) - 5\pi^2\right] \; ,\nnb\\
c^{(-1)} &=&\frac{1}{36 \,(1-x)^2} \left[-32\ln^3(1-x)+10 \pi^2 \ln(1-x) - 267\,\zeta_3 \right] \; ,\nnb\\
c^{(0)} &=&\frac{1}{(1-x)^2} \left[\frac{8}{9} \ln^4(1-x) - \frac{5}{9}\pi^2\ln^2(1-x)+8 \ln(1-x)\PL{3}{x}\right. \nnb\\
            && \hspace*{50pt} \left. +\frac{65}{3} \ln(1-x) \,\zeta_3 + 4 \, {\rm Li}_2^2(x) -\frac{167\pi^4}{270}\right] \; .
\eea
The value of the coefficient function $c^{(0)}$ at $x=0$ is one of the main new results of the present article. It confirms a conjecture
given in~\cite{Bonciani:2008wf,Beneke:2008ei} which relies on the PSLQ~\cite{pslq} algorithm. However, we emphasize that all results in the present article
were derived solely by analytical steps, and \textit{no fitting} by means of PSLQ or other methods was involved.

\section{Computation of the boundary condition}\label{sec:bc}
We now turn our attention to the computation of the boundary condition at $x=0$. Although the coefficient functions $c^{(i)}$ are
regular at $x=0$, this property cannot be used in our case to pin down the integration constant from the differential equation
method. Hence the boundary condition has to be computed by other means. Here we choose the Mellin-Barnes technique~\cite{smirnov,Tausk}.
We start from the integrand in Eq.~(\ref{eq:I}), introduce five Feynman parameters to combine the propagators and subsequently integrate
over the loop momenta. At this stage we set $x=0$ and carry out two of the Feynman parameter integrations. After applying analytic
continuation formulas for hypergeometric functions~\cite{thebook} we arrive at
\bea
&&I(x=0) = -\ESGamma^2 \, \Gamma^2(1-\eps) \, \Gamma(1+2\eps) \, (m_b^2)^{-2-2\eps} \nnb \\
&& \Bigg\{ -\int\limits_0^1 \! du \, dy \, dz \, \frac{z^{-1-\eps} \, \bar z^{\, -1-\eps} \, u^{\eps} \, \bar u^{\, -1-2\eps} \, \bar
y^{\, -1-2\eps}}{(1+2\eps) \, (\bar u +u y)^{1+2 \eps}} \, {}_2 F_1 (1+2\eps,1+2\eps;2+2\eps;-\frac{y(uy+\bar u z)}{z \bar u \bar y
(uy+\bar u)})\nnb\\
&& +\int\limits_0^1 \! du \, dy \, dz \, \frac{z^{-1-\eps} \, \bar z^{\, -1-\eps} \, u^{\eps} \, \bar u^{\, -1-2\eps} \, \bar
y^{\, -1-2\eps}}{(1+2\eps) \, (1+u y)^{1+2 \eps}} \, {}_2 F_1 (1+2\eps,1+2\eps;2+2\eps;-\frac{y(uy+\bar u z)}{z \bar u \bar y
(uy+1)})\nnb\\
&&+\int\limits_0^1 \! du \, dy \, dz \, \frac{z^{-1+\eps} \, \bar z^{\, -1-\eps} \, u^{-\eps} \, y^{\, -1-4\eps}}
{2\eps \, \bar u \, (1+u y) \, (\bar z+u z)^{2\eps}} \, {}_2 F_1 (1,1;1-2\eps;-\frac{u y (\bar z+u z)}{z \bar u (uy+1)})\nnb\\
&&-\int\limits_0^1 \! du \, dy \, dz \, \frac{z^{-1+\eps} \, \bar z^{\, -1-\eps} \, u^{-\eps} \, y^{\, -1-2\eps}}
{2\eps \, \bar u \, (\bar u+u y) \, (y+z \bar u \bar y)^{2\eps}} \, {}_2 F_1 (1,1;1-2\eps;-\frac{u(y+z \bar u \bar y)}{z \bar u
(uy+\bar u)})\Bigg\} \; , \label{eq:fp1}
\eea
where $\bar u =1-u$  and similar for $y$ and $z$. Since each of the hypergeometric functions possesses a one-dimensional Mellin-Barnes
(MB) representation according to
\be
{}_2 F_1 (a,b;c;-z) = \frac{\Gamma(c)}{\Gamma(a)\Gamma(b)} \MB{}{s} \, \frac{\Gamma(-s)\Gamma(a+s)\Gamma(b+s)}{\Gamma(c+s)} \, z^s \; ,
\ee
we can decompose each of the above terms via a three-dimensional MB representation and subsequently perform the
integrations over $u$, $y$, and $z$ in terms of $\Gamma$-functions. In one of the terms we can apply Barnes' second lemma. This yields 
\bea
&&I(x=0) = -\ESGamma^2 \, \Gamma^2(1-\eps) \, \Gamma(1+2\eps) \, (m_b^2)^{-2-2\eps} \nnb \\
&& \Bigg\{\!-\!\!\MB{k_1}{w_1}\MB{k_2}{w_2}\MB{k_3}{w_3} \; \Gamma(-\eps) \Gamma(-2 \eps - w_1) \Gamma(2 \eps + w_1 + 1) \Gamma(-w_2) \Gamma(
          2 \eps + w_1 + w_2 + 1)\nnb\\
	&&\times \frac{ \Gamma(\eps + w_1 + w_2 - w_3 + 1) \Gamma(
          2 w_1 + w_2 - w_3 + 1) \Gamma(-w_3) \Gamma(
          w_3 - w_1) \Gamma(-\eps - w_1 + w_3) }{\Gamma(2 \eps + 1) \Gamma(-3 \eps - w_1) \Gamma(
        2 \eps + w_1 + 2) \Gamma(-2 \eps + w_1 + w_2 - w_3 + 1) \Gamma(-2 \eps - w_1 + w_3)} \nnb\\
	&&\times \Gamma(-4 \eps - 2 w_1 - w_2 + w_3 - 1)\nnb\\ 
&& +\MB{k_1}{w_1}\MB{k_2}{w_2}\MB{k_3}{w_3}\;\Gamma(-\eps) \Gamma(-2 \eps - w_1) \Gamma(2 \eps + w_1 + 1) \Gamma(-w_2) \Gamma(
          2 \eps + w_1 + w_2 + 1)\nnb\\
	&& \times\frac{ \Gamma(\eps + w_1 + w_2 - w_3 + 1) \Gamma(
          2 w_1 + w_2 - w_3 + 1) \Gamma(-w_3) \Gamma(
          w_3 - w_1) \Gamma(-\eps - w_1 + w_3)}{\Gamma(2 \eps + 1) \Gamma(
        2 \eps + w_1 + 2) \Gamma(-\eps + w_2 + 1) \Gamma(-2 \eps + w_1 + w_2 - w_3 + 1)}\nnb \\
&& -\MB{k_1}{w_1}\MB{k_2}{w_2} \; \Gamma(-2 \eps) \Gamma(-\eps) \Gamma(\eps - w_1) \Gamma(-w_1)\Gamma(w_1 + 1) \Gamma(-w_2) \Gamma(
        w_1 + w_2 + 1)\nnb\\
	  && \times\frac{\Gamma(-4 \eps + w_1 + w_2)\Gamma(-\eps + w_1 + w_2 + 1)\Gamma(-4\eps +2 w_1 + w_2 + 1)  }{
	   \Gamma(-2 \eps + w_1 + 1) \Gamma(-4 \eps + w_1 + w_2 + 1)\Gamma(-3 \eps + w_1 + w_2 + 1)\Gamma(-2 \eps + w_1 + w_2 + 1) }\nnb \\
&& +\MB{k_1}{w_1}\MB{k_2}{w_2}\MB{k_3}{w_3} \; \Gamma(-2 \eps) \Gamma(-\eps) \Gamma(w_1 + 1) \Gamma(-w_2) \Gamma(
          w_1 + w_2 + 1)\Gamma(\eps - w_1 + w_3)\nnb\\
	&&\times \frac{ \Gamma(-\eps + w_1 + w_2 + 1) \Gamma(-4 \eps + w_1 + w_2 - 
            w_3) \Gamma(-w_3) \Gamma(w_3 + 1)  \Gamma(
          2 \eps - w_1 + w_3)}{\Gamma(
        2 \eps - w_1) \Gamma(-2 \eps + w_1 + 1) \Gamma(-4 \eps + w_1 + w_2 + 1) \Gamma(
        w_3 - w_1) \Gamma(-\eps - w_1 + w_3)}\nnb\\
	&&\times \; \Gamma(-2 w_1 - w_2 + w_3 - 1) \Gamma(-w_1)\Bigg\} \; .\label{eq:MB}
\eea
We preserve the order of the terms throughout this section.
In the above equation~(\ref{eq:MB}) the contour integrals in the complex 
plane can be chosen along straight lines parallel to the imaginary axis,
i.e.\ the real parts $k_i$ along the curves are constant. According to Refs.~\cite{Tausk,Alejandro,smirnovbook}, these real parts,
together with the parameter $\eps$, must be chosen in such a way as to have positive arguments in all occurring $\Gamma$-functions in
order to separate left poles of $\Gamma$-functions from right ones.
Therefore the four terms in Eq.~(\ref{eq:MB}) are regulated, respectively, for~\cite{czakonMB}
\be
\displaystyle \eps = - \frac{5}{128} \, , \quad  \displaystyle k_1 = - \frac{5}{8} \, , \quad
\displaystyle k_2= - \frac{1}{4} \, , \quad \displaystyle k_3= - \frac{9}{16} \, ,
\ee
\be
\displaystyle \eps = - \frac{7}{64}  \, , \quad \displaystyle k_1 = - \frac{9}{32}  \, , \quad
\displaystyle k_2= - \frac{1}{4} \, , \quad  \displaystyle k_3= - \frac{3}{16} \, ,
\ee
\be
\displaystyle \eps = - \frac{17}{80}  \, , \quad \displaystyle k_1 = - \frac{2}{5}  \, , \quad
\displaystyle k_2= - \frac{3}{40}  \, ,
\ee
\be
\displaystyle \eps = - \frac{185}{768}  \, , \quad \displaystyle k_1 = - \frac{229}{384}  \, , \quad
\displaystyle k_2= - \frac{25}{128} \, , \quad  \displaystyle k_3= - \frac{11}{192} \, .
\ee
We perform the analytic continuation to $\eps=0$ with the package MB~\cite{czakonMB}, which is also used for numerical
cross-checks. We then apply Barnes' lemmas and the theorem of residues on the multiple Mellin-Barnes integrals, and insert integral
representations of hypergeometric functions as well as $\psi$-functions and Euler's $B$-function where appropriate. The final result
reads
\bea
I(x=0) &=& -\ESGamma^2 \, (m_b^2)^{-2-2\eps} \nnb \\
&& \Bigg\{ \left(-\frac{\pi^2}{24\eps^2} + \frac{13\zeta_3}{4\eps}-\frac{31\pi^4}{360}\right)\nnb\\
&& +\left(- \frac{13\zeta_3}{4\eps}-\frac{\pi^4}{5}+24 \, \PL{4}{\textstyle\frac{1}{2}\displaystyle}+\ln^4 2-\pi^2\ln^2 2+21\zeta_3\ln 2\right)\nnb\\
&& +\left(-\frac{1}{24\eps^4}-\frac{\pi^2}{8\eps^2} -\frac{13\zeta_3}{4\eps}-\frac{19\pi^4}{360}-24 \, \PL{4}{\textstyle\frac{1}{2}\displaystyle}-\ln^4 2+\pi^2\ln^2 2-21\zeta_3\ln 2\right)\nnb\\
&& +\left(\frac{1}{8\eps^4}+\frac{7\pi^2}{72\eps^2} -\frac{25\zeta_3}{6\eps}-\frac{151\pi^4}{540}\right)+ {\cal O}(\eps)\Bigg\} \nnb\\
&=& -\ESGamma^2 \, (m_b^2)^{-2-2\eps}\Bigg\{\frac{1}{12\eps^4}-\frac{5\pi^2}{72\eps^2} -\frac{89\zeta_3}{12\eps}-\frac{167\pi^4}{270}+
{\cal O}(\eps)\Bigg\} \; . \label{eq:resultBC}
\eea
The calculation of all but the second term in the above equation is more or less straightforward. The second term, however, requires
more effort. In this term, at the stage when all integrations have been carried out, we were still left with terms that contain HPLs of
the sixth root of unity. Only after the application of the relations given in section~\ref{sec:hpl} we were able to write the result in
the form~(\ref{eq:resultBC}).

\section{Relations between HPLs of the sixth root of unity}\label{sec:hpl}

Higher transcendental functions that have as arguments powers of the sixth root of unity have been investigated at several places in the
literature~\cite{Broadhurst:1998rz,Fleischer:1999mp,Kalmykov:2000qe,Davydychev:2000na,Davydychev:2003mv,Kalmykov:2005hb}. We introduce
the notation
\be
\lambda_{\pm} = \frac{1}{2}\pm \frac{i}{2}\sqrt{3} = \displaystyle e^{\pm i \frac{\pi}{3}}
\ee
from which follows immediately $\lp + \lm = \lp \lm =1$. The relations among harmonic polylogarithms of weight four that we found during
the course of the calculation of the second term in Eq.~(\ref{eq:MB}) read
\bea\label{eq:HPL1}
-5 \, {\rm HPL}(\{-3,1\},\lp) - 23 \, {\rm HPL}(\{3,-1\},\lp) - 12 \, {\rm HPL}(\{2,-1,-1\},\lp) && \nnb\\
+6 \, {\rm HPL}(\{2,-1,1\},\lp) + 6 \, {\rm HPL}(\{2,1,-1\},\lp)&& \nnb\\
-6 \PL{4}{\textstyle\frac{1}{2}\displaystyle}-\frac{1}{4}\ln^4 2+\pi^2\ln^2 2-\frac{1493\,\pi^4}{38880} +c.\,c. &&=0 \, ,
\eea
\bea\label{eq:HPL2}
10 \, {\rm HPL}(\{-3,1\},\lp) + 73 \, {\rm HPL}(\{3,-1\},\lp) - 18 \, {\rm HPL}(\{-2,-1,1\},\lp) && \nnb\\
-18 \, {\rm HPL}(\{-2,1,-1\},\lp)+24 \, {\rm HPL}(\{2,-1,-1\},\lp) + 21 \, {\rm HPL}(\{-2,1,1\},\lp)&& \nnb\\
+21 \PL{4}{\textstyle\frac{1}{2}\displaystyle}+\frac{7}{8}\ln^4 2-\frac{7}{2} \pi^2\ln^2 2+\frac{5083\,\pi^4}{38880} +c.\,c. &&=0 \,
,\nnb \\
\eea
\bea\label{eq:HPL3}
-5 \, {\rm HPL}(\{-3,1\},\lp) + 9 \, {\rm HPL}(\{3,-1\},\lp) - 6 \, {\rm HPL}(\{-2,1,-1\},\lp) && \nnb\\
+3 \, {\rm HPL}(\{-2,1,1\},\lp) + 4 \, {\rm HPL}(\{2,1,-1\},\lp)&& \nnb\\
+ \PL{4}{\textstyle\frac{1}{2}\displaystyle}+\frac{1}{24}\ln^4 2-\frac{5}{6}\pi^2\ln^2 2+\frac{791\,\pi^4}{19440} +c.\,c. &&=0 \, .
\eea
Relations similar to the above ones were investigated systematically in~\cite{Broadhurst:1998rz}. We find it nevertheless
instructive to give a few comments on their derivation.

The first
relation, Eq.~(\ref{eq:HPL1}), can be derived quite easily with the package {\tt HPL}~\cite{HPL,HPL2} by applying argument
transformations to harmonic polylogarithms and exploiting the fact that $\lp = 1/\lm$ and $\lp=1-\lm$.

The second one we obtained in the
following way. In the second term in Eq.~(\ref{eq:MB}), after the analytic continuation to $\epsilon=0$, we cannot proceed until the
end by merely applying Barnes' Lemmas and corollaries thereof. One way of performing the calculation in this case is to introduce new
Feynman parameters as integral representations of higher transcendental functions (Euler's $B$-function$, \psi$-functions,
hypergeometric functions) in order to be able to perform all MB integrations.
Subsequently, one has to integrate over the newly introduced parameters. In our case, there are three such Feynman parameters to be
introduced. Note that the three newly introduced Feynman parameters are different and independent from the three-fold integration in~(\ref{eq:fp1}) which lead
towards~(\ref{eq:MB})\footnote{We tried indeed to compute the second term directly from~(\ref{eq:fp1}), without introducting MB
integrations. We succeeded to compute the simple pole, but not the finite part.}. At the stage where only
one of these new Feynman parameters, say $t$, is left, we computed the occurring integrand twice by making two different choices of
integrating by parts certain terms. The two results are then equated.

The third relation, alike the second, also
comes from the second term in~(\ref{eq:MB}), and also from the stage in where only one new Feynman parameter $t$ is left in the
integrand. Here we consider the four integrals
\bea
J_1 & \equiv & \int\limits_0^1 \! dt \; \frac{\ln(1-t)\ln^2(1-t+t^2)}{t} \, , \nnb \\
J_2 & \equiv & \int\limits_0^1 \! dt \; \frac{\PL{2}{t-t^2} \ln(1-t+t^2)}{t} \, , \nnb\\
J_3 & \equiv & \int\limits_0^1 \! dt \; \frac{\PL{2}{\displaystyle\frac{t}{1-t+t^2}} \ln(1-t+t^2)}{t} \, , \nnb
\eea
\bea\label{eq:polyint1}
J_4 & \equiv & \int\limits_0^1 \! dt \; \frac{\PL{2}{\displaystyle\frac{t^2}{1-t+t^2}} \ln(1-t+t^2)}{t} \, .
\eea
We compute each of them separately and subsequently equate them according to
\be\label{eq:PL0}
\PL{2}{\displaystyle\frac{t^2}{1-t+t^2}} + \ln(1-t)\ln(1-t+t^2) + \PL{2}{t-t^2} - \PL{2}{\displaystyle\frac{t}{1-t+t^2}} = 0 \; , \quad
0<t<1 \; .
\ee
Let us briefly outline the computation of one of the above integrals, namely $J_3$. We first observe that $\ln(1-t+t^2) = \ln(1-\lp
t)+\ln(1-\lm t)$. Therefore, we only need to compute the integral
\be
\int\limits_0^1 \! dt \; \frac{\PL{2}{\displaystyle\frac{t}{1-t+t^2}} \ln(1-\lp t)}{t} \; ,
\ee
which we consider as a function of the complex variable $\lp$. We first integrate by parts in order to be able to differentiate the
polylogarithm. We then write $1-t+t^2 = (1-\lp t) (1-t/\lp)$. In the terms containing $\ln(1-t/\lp)$ we subdivide the interval from
$0\ldots \lp$ and $\lp\ldots 1$ due to the branch cut of the logarithm. We then take the derivative w.r.t.\ $\lp$, which allows us to
perform the integration over $t$. We finally integrate the result w.r.t.\ $\lp$. The determination of the remaining additive constant is
trivial. The computation of the other $J_i$ proceeds along the same lines. 

Let us conclude this section by giving a few more relations between polylogarithms which turned out to be useful during the calculation.
See also~\cite{Davydychev:2000na,Schroder:2005va}.
\be
\PL{3}{1+\lp} = \frac{13}{18} \, \zeta_3 + \frac{\pi^2}{18} \ln(3) +i \left(\frac{\pi}{10} \, \ln^2(3) +
\frac{19\pi^3}{405}-\frac{\Phi(-\frac{1}{3},3,\frac{1}{2})}{10\sqrt{3}}\right) \; ,
\ee
\bea
{\rm HPL}(\{-2,2\},\lp) + {\rm HPL}(\{-2,2\},\lm) &=& \frac{109\pi^4}{19440}-\frac{\pi^2\ln^2(3)}{150} -
\frac{\Phi(-\frac{1}{3},2,\frac{1}{2})^2}{50} \nnb \\
&& - \frac{\pi\ln(3) \Phi(-\frac{1}{3},2,\frac{1}{2})}{25\sqrt{3}} \; ,\\
\Phi\!\left(z,s,a\right) &=& \sum\limits_{k=0}^{\infty} \frac{z^k}{[(k+a)^2]^{s/2}} \; ,\\
\Phi(\textstyle-\frac{1}{3},2,\frac{1}{2}) &=& 4 \, \sqrt{3} \; {\rm
Im}\!\!\left[\PL{2}{\textstyle\frac{i}{\sqrt{3}}}\displaystyle\right] 
= -\frac{\pi\ln(3)}{\sqrt{3}}+\frac{10}{\sqrt{3}} \; {\rm Cl}_2\!\!\left(\frac{\pi}{3}\right) \; , \nnb \\
&&\\
\Phi(\textstyle-\frac{1}{3},3,\frac{1}{2}) &=& 8 \, \sqrt{3} \; {\rm
Im}\!\!\left[\PL{3}{\textstyle\frac{i}{\sqrt{3}}}\displaystyle\right] \; .
\eea
All these relations also allow us to write the integrals $J_i$ in the following simple form
\bea
J_1 &=& \frac{271\pi^4}{4860}-\frac{2\pi^2\ln^2(3)}{75} +
\frac{3\,\Phi(-\frac{1}{3},2,\frac{1}{2})^2}{25} - \frac{4\pi\ln(3) \Phi(-\frac{1}{3},2,\frac{1}{2})}{25\sqrt{3}}
-\frac{2\pi\,\Phi(-\frac{1}{3},3,\frac{1}{2})}{5\sqrt{3}}\; , \nnb \\
\eea
\bea
J_2 &=& -\frac{139\pi^4}{1944}+\frac{\pi^2\ln^2(3)}{25} -
\frac{2\,\Phi(-\frac{1}{3},2,\frac{1}{2})^2}{25} + \frac{6\pi\ln(3) \Phi(-\frac{1}{3},2,\frac{1}{2})}{25\sqrt{3}}
+\frac{2\pi\,\Phi(-\frac{1}{3},3,\frac{1}{2})}{5\sqrt{3}}\; , \nnb \\
     && \\
J_3 &=& -\frac{989\pi^4}{9720}+\frac{\pi^2\ln^2(3)}{15} 
 + \frac{2\pi\ln(3) \Phi(-\frac{1}{3},2,\frac{1}{2})}{5\sqrt{3}}
+\frac{2\pi\,\Phi(-\frac{1}{3},3,\frac{1}{2})}{5\sqrt{3}}\; , \\
J_4 &=&  -\frac{209\pi^4}{2430}+\frac{4\pi^2\ln^2(3)}{75} -
\frac{\Phi(-\frac{1}{3},2,\frac{1}{2})^2}{25} + \frac{8\pi\ln(3) \Phi(-\frac{1}{3},2,\frac{1}{2})}{25\sqrt{3}}
+\frac{2\pi\,\Phi(-\frac{1}{3},3,\frac{1}{2})}{5\sqrt{3}}\; . \nnb \\
\eea
It is almost needless to say that also in this section all formulas were derived by purely analytical steps.

\section{Conclusion}\label{sec:conc}

We presented the result of the two-loop crossed vertex master-integral with two massive lines. Our result represents the last missing
piece in the fully analytic computation of the two-loop matching coefficients in heavy-to-light decays. The result was obtained by means
of the differential equation technique. The boundary condition at $x=0$ cannot be inferred from the condition that the integral be
regular at $x=0$. Hence we computed it explicitly from a three-dimensional Mellin-Barnes expression.

We emphasize again that all our results have been obtained by purely analytical steps, and \textit{no fitting} of numerical values to
transcendental constants by means of PSLQ~\cite{pslq} or other methods has been used.

\acknowledgments

I would like to thank D.~Ma{\^i}tre and P.~Mastrolia for useful discussions.
I acknowledge hospitality from the CERN theory group, where most of this work was performed. 
This work was supported by DFG, SFB/TR 9 ``Computergest\"{u}tzte Theoretische Teilchenphysik'', and by the German Federal Ministry of
Education and Research (BMBF).

\end{document}